\newcommand{\di}{\partial}
\newcommand{\kv}{{\boldmath{\mbox{${\scriptstyle k}$}}}}
\newlength{\parht}
\newlength{\pardp}
\documentstyle[eqsecnum,epsf,preprint,aps,aps10]{revtex}
\begin{document}
\tighten
\preprint{WISC-MILW-99-TH-02}
\draft
\title{Quantum Flux from a Moving Spherical Mirror}
\author{Warren G. Anderson}
\address{Department of Physics, \\University of Wisconsin -- Milwaukee,\\
P.O. Box 413,\\ Milwaukee, WI, USA 53201}
\author{Werner Israel}
\address{Canadian Institute for Advanced Research Cosmology Program, \\
Department of Physics and Astronomy, \\University of Victoria,\\
Victoria, BC, Canada V8W 3P6}
\maketitle
\begin{abstract}
We calculate the flux from a spherical mirror which is expanding or
contracting with nearly uniform acceleration. The flux at
an exterior point (which could in principle be a functional of the mirror's
past history) is actually found to be a local function, depending on the 
first and second time derivatives of acceleration at the retarded time.
\end{abstract}


\section{Introduction} \label{sec:Intro}
Some of the most remarkable predictions of quantum field theory arise from 
zero point fluctuations of quantum states. One of the best known examples of
this is the Unruh effect \cite{Fulling:73,Davies:75,Unruh:76}, in
which an accelerating detector measures the zero point fluctuations of the
inertial (Minkowski) vacuum state and finds they have a thermal spectrum. 
Another is the Casimir effect \cite{Casimir:48}, in which the walls (or 
boundaries) of a box experience a net force due to the difference between the
zero point fluctuations of the states inside the box and out.

It seems reasonable, then, to expect accelerating boundaries to produce
interesting effects, and indeed they do. Following studies of the Casimir
effect between moving mirrors in 1+1 dimensions by Moore \cite{Moore:70},
DeWitt pointed out that the single moving mirror problem would be interesting
and could be solved exactly in (1+1)D\cite{DeWitt:74} . Fulling
\cite{Fulling:73} and DeWitt\cite{DeWitt:75} have shown that a {\em uniformly}
accelerating mirror will indeed alter the quantum state in the vicinity of the
mirror. However, a far more interesting result was obtained by Davies and
Fulling\cite{FullingDavies:76,DaviesFulling:77} for a mirror
experiencing {\em non-uniform} acceleration in (1+1)D. Such a mirror actually
emits fluxes of quantum radiation, as though the mirror were knocking zero
point quanta out of the vacuum and off to infinity. More precisely, they found
that a mirror with 2-velocity $u^\mu = d x^\mu/d\tau$ ($u\cdot u = -1$)
and acceleration $a^\mu$ emits a flux 
\begin{equation}
   \frac{dE}{d\tau} = - \langle T_{\mu\nu}\rangle u^\mu n^\nu = 
   -\frac{\hbar}{12\pi} \frac{d}{d\tau}~(a\cdot n),
\label{eq:1Dflux}
\end{equation}
in the direction of a unit spatial vector $n^\nu$ orthogonal to $u^\mu$. This
holds for either choice (``left'' or ``right'') of $n^\mu$.  Thus, a mirror
whose acceleration is increasing (algebraically) toward the right will emit
a stream of negative energy to the right and a numerically equal positive
stream to the left.

The implications of this result are intriguing. Davies\cite{Davies:72} and
Ford\cite{Ford:78} first raised the possibility that the negative-energy flux
from a moving mirror could be used to cool a hot body and thus violate the
second law of thermodynamics in a quantum context, and this paradox was
further discussed by Deutsch, Ottewill and Sciama\cite{Deutschetal:82}.
Limitations on the extent of such violations in flat spacetime (``quantum
interest'') have been formulated by Ford and Roman\cite{FordRoman:99} and
others (\cite{Flanagan:97}, \cite{Pretorius:99}, {\em etc.}).

More recently, Anderson\cite{Anderson:94} has used this result in the context
of the Geroch gedankenexperiment\cite{Bekenstein:81}. In this experiment, a
box with mirrored walls is filled with radiation and lowered adiabatically 
toward a black hole. Unruh and Wald\cite{UnruhWald:82} have shown that such a 
box is subject to a buoyancy force and will eventually reach a floating point 
above the black hole. Anderson has examined this further and shown that the
ground state inside the box is the Boulware state, whose energy (which becomes
increasingly negative as the box descends) is fed by Davies-Fulling fluxes 
from the reflecting walls. This accounts for the buoyancy felt by the box.

The volume of literature on moving mirrors is impressive, but it bears noting 
that all the results mentioned above are obtained in 1+1
dimensions. Indeed, if one includes the result for moving mirrors in curved
space-times obtained by Ottewill and Takagi\cite{OttewillTakagi:88} with those
reviewed above, the (1+1)D theory of moving mirrors can be considered
essentially complete. This is due largely to the conformal properties of
quantum field equations in (1+1)D, which allow boundaries, and even
space-time itself, to be flattened, thereby enabling one to obtain results for
complicated geometries from those for much simpler geometries.

This is not the case in 3+1 dimensions, where only partial results are
available. The case of constant acceleration has been solved for both
plane\cite{CandelasDeutsch:77,CandelasDeutsch:78} and
spherical\cite{FrolovSerebriany:79} mirror geometries. Ford and
Vilenkin\cite{FordVilenkin:82} have extended the plane mirror result to
include non-constant acceleration for the case when the acceleration and its
derivatives are small. More recently, Hadasz {\em et al.}\cite{HadaszEtc:98}
have considered arbitrary (radial) motion of a spherical
mirror, but have restricted their attention to the ``S-wave approximation''
where only spherically symmetric modes are considered. Because of this
restriction, their result can be related to the 1+1 dimensional results of
Davies and Fulling\cite{FullingDavies:76,DaviesFulling:77}.

Consideration of quasi-stationary processes (e.g. slow descent of a mirror in
a strong gravitational field) requires knowledge of the flux emitted by a
mirror whose acceleration is changing slowly, though it may be large. Our
objective in this paper is to derive an interesting and relatively simple
result of this type in 3+1 dimensions. The central tool is the use of a
Green's function perturbation technique.  Evaluating the perturbation is much
more manageable if the Green's functions for the unperturbed problem are
available in closed form. This is actually the case for a uniformly
accelerated spherical mirror, as shown by Frolov and
Serebriany\cite{FrolovSerebriany:79}. The mirror's history is then a
three-dimensional pseudo-sphere of radius $b$, say, and the unperturbed
problem is just the Minkowski-signature analogue of finding the
four-dimensional electrostatic potential of a point charge in the presence of
an earthed conducting 3-sphere of radius $b$. This is easily solved by the
method of images.

Our objective in this paper is to solve the perturbed Frolov-Serebriany
problem, {\em i.e.} to examine the effect of small spherically symmetric
non-uniformities in the acceleration of a spherical mirror.

It is well to stress at the outset that the solution for a plane mirror cannot
be derived from ours by a straightforward limiting process. The single
parameter $b$, whose reciprocal gives the unperturbed acceleration, also gives
the minimum radius attained by the mirror as seen from its center. Thus, the
plane limit $b\rightarrow\infty$ is inseparable from small acceleration.
(There are reasons to expect the planar case to be considerably more
complicated. Formally, the Wightman Green's function is now an infinite sum of
McDonald (Bessel) functions. Geometrically, any light ray reflected
non-orthogonally of a uniformly accelerated plane mirror will re-encounter the
mirror an infinite number of times; in the spherical case there is just one
encounter.)

Also, we concern ourselves only with calculating the outward flux, which we
expect to be the most interesting stress-energy component. In fact, it turns
out to be somewhat more interesting than one might expect. We find that it
has a remarkable property; although it could, in principle, depend on
the entire retarded history of the mirror, to first order in the mirror
perturbation it depends only the behavior of the mirror at the most recent 
retarded time, {\em i.e.} it is local.

This article is organized as follows: in Section \ref{sec:UniAcc} we review
the Frolov-Serebriany result for a mirror expanding with uniform acceleration.
In Section \ref{sec:Perturb} we present our Green's function perturbation
scheme, and in Section \ref{sec:Eval} we use it to evaluate the corrections to
the Frolov-Serebriany Green's function, with some of the more cumbersome
details relegated to Appendix \ref{sec:AppA}. Section \ref{sec:Flux} is
concerned with calculating the quantum flux from these perturbations, with
details again left to Appendices (\ref{sec:AppB} and \ref{sec:AppC}).  
Finally, in Section \ref{sec:Conclusion} we offer some concluding remarks.

\section{Uniformly accelerating spherical mirror}\label{sec:UniAcc}
In the case where the mirror's acceleration is uniform, the Green's functions
for the massless fields can be obtained in simple closed form by the method of
images, as noted by Frolov and Serebriany\cite{FrolovSerebriany:79}. In this
Section we shall briefly review these results.

Consider first the static potential due to a point charge $q'$ in Euclidean
4-space at a distance $R'$ from the center of an earthed conducting 3-sphere 
of radius $b$. The Dirichlet boundary condition can be reproduced by
introducing a co-radial image charge $q''=-(b/R')^2q'$ at radius $R''=b^2/R'$.

In the Lorentzian analogue of this problem, we are concerned with Green's
functions for the wave equation $\Box \varphi = 0$ in Minkowski space-time,
with Dirichlet boundary conditions on the pseudo-sphere ({\em i.e.} the
time-like hyperboloid of one sheet) $R=b$, where now
\begin{equation}
   R^2 \equiv \eta_{\mu\nu}x^\mu x^\nu=x^2+y^2+z^2-t^2=\rho^2+z^2-t^2,
\label{eq:rad}
\end{equation}
in a self-evident notation.

The pseudo-sphere $R=b$ represents the history of a spherical mirror of radius
$b$ (constant as measured in its instantaneous rest frame), whose center is
fixed at the spatial origin $x=y=z=0$ and which moves with uniform
acceleration $a=b^{-1}$. 

The image construction gives for the retarded Green's function
$G_{ret}(x,x')$, satisfying 
\begin{equation}
   \Box G_{ret}(x,x')=-\delta^4(x,x'),
\label{eq:reteq}
\end{equation}
the expression 
\begin{equation}
   G_{ret}(x,x')=\frac{1}{2\pi}~\theta(t-t')\left\{\delta[(xx')^2]-\left(
      \frac{b}{R'}\right)^2\delta[(xx'')^2]\right\}.
\label{eq:Gretconst}
\end{equation}
Here $\theta$ is the unit step (Heaviside) function, $\delta$ is the Dirac
distribution, $(xx')^2$ is the squared Minkowski interval
\begin{equation}
   (xx')^2\equiv\eta_{\mu\nu}(x^\mu-{x'}^\mu)(x^\nu-{x'}^\nu),
\label{eq:Minkint}
\end{equation}
and the image source is located at
\begin{equation}
   {x''}^\mu = \left(\frac{b}{R'}\right)^2{x'}^\mu.
\label{eq:imagepoint}
\end{equation}

Similarly, the Wightman function
\begin{equation}
   W(x,x')=\langle0|\varphi(x)\varphi(x')|0\rangle
\label{eq:Wdef}
\end{equation}
for a massless scalar field takes the form 
\begin{equation}
   W(x,x')=\frac{1}{4\pi^2}\left\{\frac{1}{(xx')^2+i(t-t')\epsilon} -\left(
      \frac{b}{R'}\right)^2\frac{1}{(xx'')^2+i(t-t'')\epsilon}\right\},
\label{eq:Wconst}
\end{equation}
with $\epsilon \rightarrow +0$.

\section{Nearly uniform acceleration: perturbing the boundary}
\label{sec:Perturb}
The corresponding Green's functions for a spherical mirror whose acceleration
is slightly non-uniform can be derived from the preceding results by
superposing the effect of a small perturbation on the history of the mirror,
{\em i.e.}, the time-like 3-space $\Sigma$ on which Dirichlet boundary
conditions are imposed.

Consider generally the problem of solving 
\begin{equation}
   \Box \Phi = 0, \hspace{1cm} \Phi=0 ~~\mbox{on}~~ \Sigma.
\label{eq:BVP}
\end{equation}
Suppose that $\Sigma$ is a small perturbation of a simpler time-like 3-space
$\Sigma_0$, obtained by displacing $\Sigma_0$ a distance $\delta n(x)$ along
its outward normal $n$, and that we know the solution $\Phi_0$ of the problem
\begin{equation}
   \Box \Phi_0 = 0, \hspace{1cm} \Phi=0 ~~\mbox{on}~~ \Sigma_0,
\label{eq:unpertBVP}
\end{equation}
with the same initial boundary conditions.

Then (\ref{eq:BVP}) can be reformulated as a problem with boundary conditions
specified on the unperturbed boundary $\Sigma_0$:
\begin{equation}
   \Box \Phi = 0, \hspace{1cm} \Phi=-\frac{\di\Phi_0}{\di n}~ \delta n(x)
      ~~\mbox{ on }~~ \Sigma_0.
\label{eq:pertBVP}
\end{equation}

The causal solution for the perturbation $\delta \Phi \equiv \Phi - \Phi_0$
follows from Green's identity:
\begin{equation}
   \delta \Phi(x') = \int_{\Sigma_0} d\Sigma~\frac{\di G_{ret}(x',x)}{\di n}
      ~\delta n(x)~\frac{\di}{\di n} \Phi_0(x),
\label{eq:fieldpert}
\end{equation}
where $G_{ret}$ is the retarded Green's function for the unperturbed boundary
value problem (\ref{eq:unpertBVP}) and $x$ is in $\Sigma_0$. Equation
(\ref{eq:fieldpert}) defines a linear operation applied to $\Phi_0$, which we
shall write for brevity as
\begin{equation}
   \delta \Phi(x') = L_{\delta n}(x')~\Phi_0(\cdot).
\label{eq:pertop}
\end{equation}

To obtain the effect of the perturbation on the Wightman function
(\ref{eq:Wdef}), we note that it can be written as a mode sum
\begin{equation}
   W(x,x') = \int \frac{d^3 k}{(2\pi)^3} ~f_\kv (x) ~\overline{f_\kv}(x'),
\label{eq:Wsum}
\end{equation}
where $\{ f_\kv(x)\}$ is a complete set of initially positive frequency
solutions of (\ref{eq:BVP}) and the bar denotes complex conjugation. Applying
(\ref{eq:pertop}) to each mode separately and summing the results yields
\begin{equation}
   \delta W(x,x') = L_{\delta n}(x)W_0(\cdot,x')+L_{\delta n}(x')W_0(x,\cdot),
\label{eq:modepert}
\end{equation}
where $W_0$ denotes the unperturbed Wightman function given by
(\ref{eq:Wconst}).

\section{Wightman function for nearly uniform acceleration}\label{sec:Eval}
Evaluation of the expression (\ref{eq:modepert}) for $\delta W$ in the case
where the mirror's acceleration departs slightly from uniformity is somewhat
lengthy but straightforward. Here, we outline the results of this calculation
reserving the technical details for Appendix \ref{sec:AppA}. 

It will be assumed that the mirror remains spherical as viewed by an observer
at its center $x=y=z=0$. Then its history is
\begin{equation}
   R=b(1+f(t)), \hspace{1cm} |f(t)| \ll 1
\label{eq:boundpert}
\end{equation} 
in terms of this observer's time $t$, where $f(t)$ is arbitrary but small. 
The corresponding advanced and retarded times are written
\begin{equation}
   v=t+r,\hspace{1cm}u=t-r,\hspace{1cm}r=\sqrt{x^2+y^2+z^2}.
\label{eq:uvdef}
\end{equation}

The radial energy flux measured by a stationary observer outside the mirror is 
\begin{equation}
   F= T_{uu}-T_{vv}.
\label{eq:statflux}
\end{equation}
Each of these terms can be dealt with by a similar procedure. Let us consider
$T_{uu}$; it is evident from (\ref{eq:Wdef}) that its expectation value at an
event $p_1$ outside the mirror is
\begin{equation}
   \langle T_{uu}(p_1)\rangle = \left[ \di_{u_1} \di_{u_2} W(p_1,p_2)
      \right]_{p_2=p_1}
\label{eq:fluxdef}
\end{equation}
for a minimally coupled massless scalar field, and a similar but more
complicated expression for conformal coupling (see (\ref{eq:quantSETconf})
below). Note that regularization and symmetrization are not needed to evaluate
this component of the stress-energy.

We can now present the results for $\delta W$.  The spherical symmetry of
(\ref{eq:boundpert}) allows us to take both $p_1$ and $p_2$ in the $z-t$ plane
($x=y=0$).  Calculation shows (see Appendix \ref{sec:AppA}) that the first
term of (\ref{eq:modepert}) contributes
\begin{equation}
   \frac{8\pi^2\zeta^2}{(R_1^2-b^2)(R_2^2-b^2)} L_{\delta n}(p_1)W(\cdot,p_2)
      = \frac{z_1z_2}{\zeta} \int_{-\infty}^{t_1^*} dt \frac{f(t)}{(t-t_0)^3}
      -\frac{z_1z_1^*}{(R_1^2-b^2)(t_1^*-t_0)^2}
\label{eq:Wterm1}
\end{equation}
where $\zeta$ and $t_0$ are defined by
\begin{equation}
   \zeta=z_1t_2-z_2t_1, ~~~~~\zeta t_0 = \frac{1}{2}(z_1-z_2)(z_1z_2-b^2)
      +(z_1t_2^2-z_2t_1^2),
\label{eq:zetadef}
\end{equation}
and $p_i^*$ (with coordinates $x_i^*,t_i^*$) is the event nearest to $p_i$
($i=1,2$) at which the past light-cone of $p_i$ intersects the (unperturbed)
mirror (see Fig. 1).
\begin{figure}[hbtp]
\begin{center}
\leavevmode \epsfbox{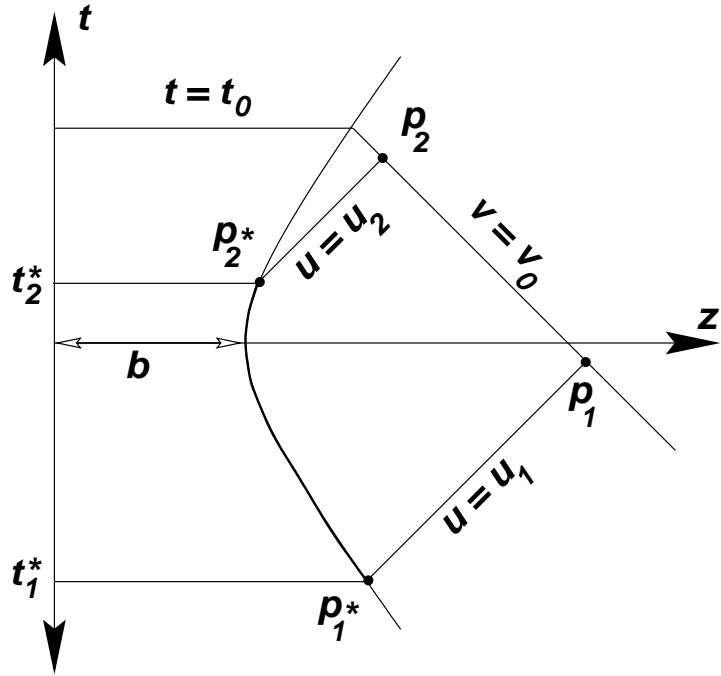}
\end{center}
\caption{The $z-t$ plane. The mirror profile is the hyperbola. $p_1$ and $p_2$
are the two events at which the Wightman function is to be evaluated and $v_0$
is their common advanced time.  $p_1^*$ and $p_2^*$ are the intersections of
their past light-cones with the mirror in the $z-t$ plane.  
\label{fig:1}}
\end{figure}

The contribution of the second term in (\ref{eq:modepert}) is obtained by
interchanging $p_1$ and $p_2$ in (\ref{eq:Wterm1}). Because $\zeta$ is an odd
function of $p_1,~p_2$ ($t_0$ is even), the term involving the integral
changes sign. Thus, the sum of the two contributions,
\begin{equation}
   -\frac{8\pi^2\zeta^2}{(R_1^2-b^2)(R_2^2-b^2)}\delta W(p_1,p_2) = 
      \frac{z_1z_2}{\zeta} \int_{t_1^*}^{t_2^*} dt \frac{f(t)}{(t-t_0)^3}
      +\sum_{i=1}^2\frac{z_iz_i^*}{(R_i^2-b^2)(t_i^*-t_0)^2}f(t_i^*),
\label{eq:deltaW}
\end{equation}
depends only on the mirror's history between the retarded times $t_1^*$ and
$t_2^*$. Particle and anti-particle modes interfere destructively in the case
of a spherical mirror to eliminate the effects of the past history and 
produce (when we take the coincidence limit $p_2 \rightarrow p_1$) a
purely local expression for the stress-energy.

To perform the partial derivatives $\di_{u_1}$ and $\di_{u_2}$ in the
expression (\ref{eq:fluxdef}) for~~$T_{uu}$, we require mutual independence of
the $u$-coordinates of the events $p_1$ and $p_2$. But these are the only
coordinates which need be independent. To evaluate $T_{uu}$, it suffices to
consider $\delta W(p_1,p_2)$ in a partial pre-coincidence limit $v_1=v_2$, as
in Fig. \ref{fig:1}. 

These remarks in principle apply, {\em mutatis mutandis}, also to the 
evaluation of~~$T_{vv}$, but there is a complication. In the pre-coincidence
limit $u_2\rightarrow u_1$, the points $p_1^*$ and $p_2^*$ tend to coincidence
with each other and with the point on the mirror having coordinate $t_0$, so
that the integrand in (\ref{eq:deltaW}) becomes infinite while the interval of
integration shrinks to zero. The evaluation of $T_{vv}$ is discussed further in
Appendix \ref{sec:AppC}. Here, we merely note that $T_{vv}$ makes no
radiative contribution (proportional to $r^{-2}$) to the flux
(\ref{eq:fluxdef}). This follows at once from the identity
\begin{equation}
   \di_u\left(r^2 \di_u \left(r^2T_{vv}\right)\right) =  \di_v 
      \left( r^2 \di_v \left( r^2 T_{uu}\right)\right),
\label{eq:consEq}
\end{equation}
which is a consequence of the conservation of $T_{\mu \nu}$ and the vanishing
of the trace $T^\alpha_\alpha$ for a conformal scalar field in flat space. In
(\ref{eq:consEq}), the derivative $\di_v$ increases the fall-off with
distance, but this does not hold for $\di_u$, which can operate on the
retarded displacement $f(u)$ in (\ref{eq:boundpert}). Thus, $T_{vv}$ falls off
more strongly than $T_{uu}$. The detailed calculation (Appendix
\ref{sec:AppC}) shows that $T_{vv}\sim r^{-6}$ as $r\rightarrow\infty$.

\section{Flux from a non-uniformly accelerating mirror}\label{sec:Flux}
The expectation value of the stress-energy is derivable from the partial
derivatives of the Wightman function $W(x,x')=W_0+\delta W$ in the coincidence
limit, with the unperturbed part $W_0$ given by (\ref{eq:Wconst}) and the
perturbation $\delta W$ by (\ref{eq:deltaW}). The unperturbed part becomes
singular in the limit $x' \rightarrow x$, but is easily regularized by
subtracting the value of $W_0$ in free space without the mirror, {\em i.e.},
the first term of (\ref{eq:Wconst}), leaving the second (image) term as the
sole contribution to $(W_0)_{reg}$. The perturbation $\delta W$ is regular in
the coincidence limit.

For a massless scalar field, two different stress tensors are commonly
considered: (a) the minimal stress-energy, given classically by
\begin{equation}
   \left(T_{\mu \nu}\right)_{min} = \varphi_{,\mu} \varphi_{,\nu} -
      \frac{1}{2} g_{\mu \nu} (\nabla \varphi)^2
\label{eq:classSETmin}
\end{equation} 
and quantum mechanically by
\begin{equation}
   \langle T_{\mu \nu}(x)\rangle_{min} = \left[(\di_{\mu}\di_{\nu'}-\frac{1}{2}
      g_{\mu\nu}~\di^{\alpha}\di_{\alpha'})W_{reg}(x,x')\right]_{sym~;~x'=x},
\label{eq:quantSETmin}
\end{equation}
in which ``sym'' indicates symmetrization in (x,x') and in the partial
derivatives; (b) the conformal (trace-free) stress-energy, defined classically
by
\begin{equation}
   \left(T_{\mu \nu}\right)_{\mathit conf} = \frac{2}{3}\varphi_{,\mu} 
      \varphi_{,\nu}- \frac{1}{3}\varphi\varphi_{,\mu\nu}-
      \frac{1}{6}g_{\mu \nu} (\nabla \varphi)^2
\label{eq:classSETconf}
\end{equation} 
and quantum mechanically by
\begin{equation}
   \langle T_{\mu \nu}(x)\rangle_{\mathit conf}=\frac{1}{3}\left[
      (2\di_{\mu}\di_{\nu'}- -\di_{\mu'}\di_{\nu'}
      -\frac{1}{2} g_{\mu\nu}~\di^{\alpha}\di_{\alpha'})
      W_{reg}(x,x')\right]_{sym~;~x'=x},
\label{eq:quantSETconf}
\end{equation}

We begin by reviewing the Frolov-Serebriany\cite{FrolovSerebriany:79} results
for {\em uniform} acceleration. Differentiating the regularized form of
(\ref{eq:Wconst}), we easily find
\begin{eqnarray}
   \langle T^{(0)}_{\mu \nu}(x)\rangle_{min} &=& - \frac{b^2}{\pi^2(R^2-b^2)^4}
      \left(x^{\mu}x^{\nu}-\frac{1}{2}g^{\mu\nu}R^2\right),
\label{eq:constSETmin}\\
      \langle T^{(0)}_{\mu \nu}(x)\rangle_{\mathit conf} &=& 0.
\label{eq:constSETconf}
\end{eqnarray}   
This last result is quite remarkable, because the conformal stress is not
likely to vanish for a spherical mirror at rest (it certainly does not for
electromagnetic fields \cite{Boyer:68,Davies:72,MiltonEtc:78}).
It appears that the effects of uniform acceleration exactly cancel the static
Casimir stresses.

\begin{figure}[hbtp]
\begin{center}
\leavevmode \epsfbox{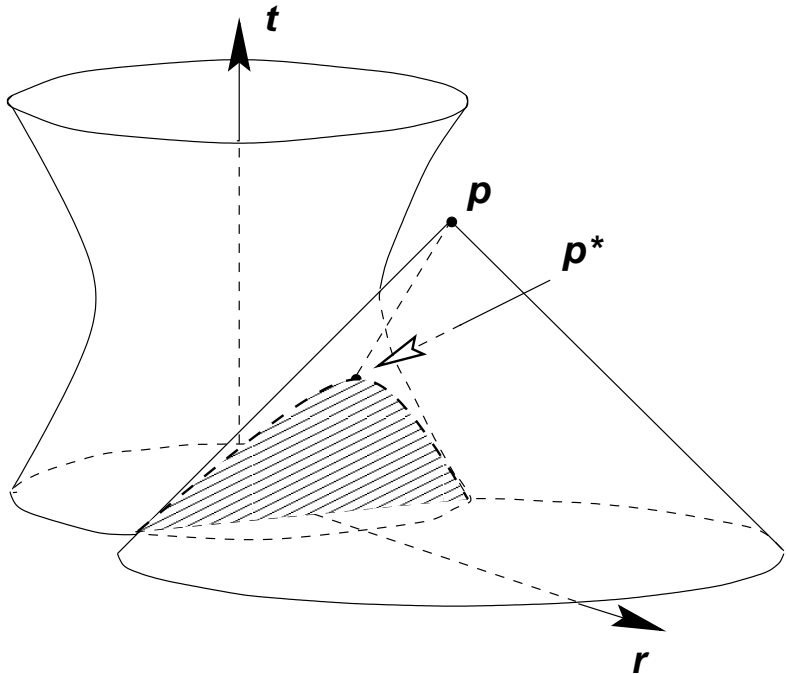}
\end{center}
\caption{Intersection of the history of a uniformly accelerating sphere
(hyperboloidal cylinder) and the past null cone of an exterior point $p$. The
intersection (which lies entirely within the shaded plane) is represented by
the bold curve. The nearest retarded point to $p$ on the mirror's worldsheet,
$p^*$, is the only point whose perturbation contributes to the flux at $p$.
This figure has been dimensionally reduced; each point represents a circle.
\label{fig:2}}
\end{figure}

The effects of non-uniform acceleration are more complicated. We shall simply
quote the result for the conformal radial out-flux $\langle T_{uu}
\rangle_{\mathit conf}$ at a point $(r,t)$ outside the mirror, leaving to
Appendix \ref{sec:AppB} an outline of the derivation:
\begin{equation}
   \langle T_{uu}(t,r) \rangle_{\mathit conf} = - \frac{1}{1440 \pi^2} 
      \frac{1}{rv} \left\{\frac{q}{m^3n}\frac{d^2\alpha}{d\chi^2} - 
      \frac{4}{m^3n^2r} (npr-q^2)\frac{d \alpha}{d \chi} +
      \frac{2 s^2}{mn^3r^2}\left[q(\alpha+f)+p\frac{df}{d\chi}\right]\right\}.
\label{eq:SETpert}
\end{equation}
The notation is as follows: the advanced and retarded times, $v$ and $u$ are
defined as in (\ref{eq:uvdef}), and we write
\begin{eqnarray}
   m&=&-\frac{u}{2},~~~n~=~\frac{1}{2v}(R^2-b^2)~=~-\frac{1}{2}(u+b^2/v),
\nonumber\\
   p&=&\frac{1}{8}(u^2-b^2),~~~q~=~\frac{1}{8}(u^2+b^2),~~~
      s~=~-\frac{1}{2}(v+b^2/v);
\label{eq:notdefs}
\end{eqnarray}
$\chi$ is the pseudo-angle along the (unperturbed) mirror trajectory ({\em
i.e.}, $\tau=b\chi$ is the mirror's proper time). Equation
(\ref{eq:boundpert}), giving the trajectory of the perturbed mirror, is now
written $R=b(1+f(\chi))$, and
\begin{equation}
   \alpha = f''(\chi) - f(\chi)
\label{eq:alphadef}
\end{equation}
is a measure of the non-uniformity of the acceleration $A$, which is given by
\begin{equation}
   A = (1+\alpha)/b.
\label{eq:acceldef}
\end{equation}
In (\ref{eq:SETpert}), $\chi$ refers to the pseudo-angle at the point $p^*$,
which is the nearest retarded point to (r,t), as in Fig. \ref{fig:2}.
The corresponding expression for the canonical flux is too long to reproduce
here, and is also deferred to Appendix B.

The limit $b \rightarrow \infty$ corresponds to a slowly accelerating, nearly
plane mirror. This was the case studied by Ford and
Vilenkin\cite{FordVilenkin:82}. It is straightforward to show that in this
limit our result (\ref{eq:SETpert}) for the conformal flux (and also our
result for the canonical flux) reduces to the expressions they give.

To obtain a more intuitive grasp of the physical meaning of the complex
expression (\ref{eq:SETpert}), we can evaluate the flux $F$ radiated at
retarded time $u=-b$ ~~-~~ {\em i.e.} when the mirror is near its minimum
radius $r_M \approx b$ ~~-~~ as measured by a stationary observer at radius
$r\gg b$ and the same retarded time. Using (\ref{eq:statflux}), taking the
appropriate limit of (\ref{eq:SETpert}), and noting that $T_{vv}$ does not
contribute to the flux to leading order (as discussed at the end of Section
\ref{sec:Eval}) we find
\begin{equation}
   F=-\frac{\hbar}{720 \pi^2}\left(\frac{R_0}{r}\right)^2\left\{A 
      \frac{d^2 A}{d \tau^2}+2A^3\left(A-\frac{1}{R_0}\right)\right\},
\label{eq:flux}
\end{equation}
where $R_0=b(1+f(0))$ is the proper radius of the mirror at the time of
emission $\chi = 0$, and we have restored Planck's constant to display the
correct dimensionality. We recall that this perturbative result is correct
to linear order in deviations from uniform acceleration $(A-1/R)$, $dA/d\tau$,
$d^2A/d\tau^2$, but the acceleration $A$ itself is arbitrary.

\section{Concluding Remarks}\label{sec:Conclusion}
Our chief interest in this paper has been in the quantum flux radiated by the
mirror and we have explicitly computed only those components of the
stress-energy tensor from which it arises. However, the remaining components
can be derived straightforwardly (though with some labour) from our expression
(\ref{eq:deltaW}) with the methods of Appendix \ref{sec:AppB}. It would be
useful to have these to round out the picture.

The relevant Green's functions for an unperturbed, uniformly accelerating
spherical mirror have a simple closed form, and this has enormously simplified
our perturbative calculation. This simplification is bought at a price: we are
limited to spherical mirrors whose acceleration $A$ and proper radius $R$ are
nearly reciprocals. We cannot decouple the plane limit $R\rightarrow \infty$
from the limit of small acceleration, and cannot disentangle curvature
(Casimir) effects from the effects of acceleration. 

It is evident that much remains to be done before we can claim anything
approaching a comprehensive understanding of the quantum dynamics of
three-dimensional mirrors. 
\acknowledgements{We are indebted to Valeri Frolov and Tom Roman for
discussions and to the latter for calling our attention to the prior work of
Ford and Vilenkin \cite{FordVilenkin:82} on slowly accelerating plane mirrors. 

This research was supported by the Canadian Institute for Advanced Research
and by NSERC of Canada.}
\appendix
\section{Evaluation of the integral (3.7) for $\delta W$} \label{sec:AppA}
To verify (\ref{eq:deltaW}), one needs to evaluate $L_{\delta
n}(p_1)~W_0(\cdot,p_2)$, where the integral operator $L_{\delta n}$ is defined
by (\ref{eq:pertop}) and $W_0$ by (\ref{eq:Wconst}). Because of the spherical
symmetry of the mirror, there is no loss of generality in taking $p_1$ to be
in the $z-t$ plane. Then the integration over the azimuthal cylindrical
coordinate $\phi$ is trivial.

The problem is essentially solved by the following lemma. Let
$F(p)=F(\rho,z,t)$ be any axisymmetric function. Then we have the identity
\begin{eqnarray}
   &&\int_{\Sigma_0} F(p)~\frac{\di}{\di n} G_{ret}(p_1,p)~d\Sigma =
      -\mbox{sgn}(R^2-b^2)~\frac{z_1^*}{z_1}~F(p_1^*)
\nonumber\\
   &&~~~~~~~~~~+\frac{1}{2}~\mbox{sgn}(z_1)
      ~\frac{R_1^2-b^2}{z_1^2}~\int_{-\infty}^{t_1^*} dt~\left[\frac{\di}{\di z}
      ~F(\rho=\sqrt{b^2+t^2-z^2},z,t)\right]_{z=Z_1(t)}
\label{eq:lemma}
\end{eqnarray} 
where $p_1^*$ is defined as in Section \ref{sec:Eval} and the linear function
$z=Z_1(t)$ is the solution of $\xi(p_1,p)=0$ where 
\begin{equation}
   \xi(p_1,p) \equiv t_1 t - z_1 z +\frac{1}{2}(R_1^2+b^2).
\label{eq:xidef}
\end{equation}
Geometrically, $z=Z_1(t)$ represents the line in the $z-t$ plane through
$p_1^*$ and orthogonal to the radius vector joining $p_1$ to the origin
$z=t=0$.

To prove (\ref{eq:lemma}), let us note that the presence of $G_{ret}$ will
effectively confine the integration to a 2-space ${\cal S}$, formed by the
intersection of the unperturbed mirror's history 
\begin{equation}
   \Sigma_0~~:~~\sigma(0,p)=\frac{1}{2}b^2
\label{eq:cylinder}
\end{equation}
with the past light-cone of $p_1$, given by 
\begin{equation}
   \sigma(p_1,p)=0,
\label{eq:cone}
\end{equation}
where $\sigma(p_1,p_2)=\frac{1}{2}(p_1~p_2)^2$ is the usual geodesic biscalar
({\em c.f.} (\ref{eq:Minkint}) and see Fig. \ref{fig:2}). Taking the
difference between (\ref{eq:cylinder}) and (\ref{eq:cone}), we see that ${\cal
S}$ can equivalently be regarded as the intersection of the mirror $\Sigma_0$
with the 3-plane $\xi(p_1,p)=0$, where
\begin{equation}
   \xi(p_1,p)=\sigma(p_1,p)-\sigma(0,p)+\frac{1}{2}b^2
\label{eq:xidef2}
\end{equation}
is the same as (\ref{eq:xidef}).

$G_{ret}$ in the integral (\ref{eq:lemma}) is a (distributional) function of
$\sigma(p_1,p)$ and $\sigma(p_1,\tilde{p})$ (see (\ref{eq:Gretconst})), where
$\tilde{p}$ is the image point of $p$. To obtain its normal derivative on the
mirror $R=b$, it is convenient to introduce four dimensional polar coordinates
for $p$,
\begin{equation}
   \rho=R \cosh \chi \sin\theta,~~~~~z=R \cosh \chi \cos\theta,~~~~~
      t=R \sinh \chi,
\label{eq:polarcoord}
\end{equation}
so that the normal derivative (``outward'' from $\Sigma_0$ as viewed from the
point $p$ outside $\Sigma_0$) corresponds to $-\di/\di R$. We easily find
\begin{eqnarray}
   2~\sigma(p_1,p)&=&R^2+2R(t_1\sinh\chi - z_1\cosh\chi\cos\theta)+R_1^2,
\label{eq:sigmacoord}\\
   \left. \frac{\di}{\di R} \sigma(p_1,p)\right|_{R=b} &=&
      \frac{1}{b}\left[\xi(p_1,p)-\frac{1}{2}(R_1^2-b^2)\right].
\label{eq:dnsigmacoord}
\end{eqnarray}
A similar calculation for the image contribution yields values for
$2\sigma(p_1,\tilde{p})$ and $\frac{\di}{\di R} \sigma(p_1,\tilde{p})$
numerically equal to (\ref{eq:sigmacoord}) and (\ref{eq:dnsigmacoord}) on the
mirror, but for the normal derivative the sign is opposite.

The element of 3-area $d\Sigma$ on the mirror
\begin{equation}
   \Sigma_0~~:~~\rho=\sqrt{b^2+t^2-z^2},
\label{eq:Sigmacoord}
\end{equation}
employing $(t,z,\phi)$ as intrinsic coordinates, is $d\Sigma = b~d\phi~dz~dt$.
Hence the integral of an axisymmetric function $H(\rho,z,t)$ takes the explicit
form
\begin{equation}
   \int_{\Sigma_0} d \Sigma~H(\rho,z,t) = 2\pi b \int_{-\infty}^\infty
      \int_{-\infty}^\infty dt~dz~\theta(b^2+t^2-z^2)~
      H(\rho=\sqrt{b^2+t^2-z^2},z,t)
\label{eq:intH}
\end{equation}
in which the step function $\theta$ takes into account that, for fixed $t$,
the range of $z$ over $\Sigma_0$ is restricted by (\ref{eq:Sigmacoord}).

The distributional factors $\delta'[2\sigma(p_1,p)]$ and
$\delta[2\sigma(p_1,p)]$, which arise from $\di G_{ret}(p_1,p)/\di n$ in the
integral over $\Sigma_0$, are handled as follows. For points $p$ restricted to
$\Sigma_0$, (\ref{eq:cylinder}) and (\ref{eq:xidef2}) show that
$\sigma(p_1,p)=\xi(p_1,p)$, given explicitly in (\ref{eq:xidef}) in terms of
the intrinsic coordinates $(t,z,\phi)$ of $\Sigma_0$. Taking the partial
derivatives with respect to $z$ tangentially along $\Sigma_0$ ({\em i.e.},
holding $\phi$ and $t$ fixed in $\xi$), we immediately find
\begin{equation}
   \frac{\di}{\di z} \delta[2\sigma(p_1,p)] = -2 zz_1 \delta'[2\sigma(p_1,p)].
\label{eq:delta'}
\end{equation}
Thus, $\delta'$ can be eliminated in favor of the tangential derivative $\di
\delta / \di z$, which can be converted through integration by parts in
(\ref{eq:intH}) to 
\begin{equation}
   \delta[2\sigma(p_1,p)] = \delta[2\xi(p_1,p)] = \frac{1}{2~|z_1|}~
      \delta(z-Z_1(t)),
\label{eq:deltatrans}
\end{equation}
by virtue of (\ref{eq:xidef}).

Putting all this together leads straightforwardly to the quoted result
(\ref{eq:lemma}).

\section{Derivation of (5.7) for the conformal flux}\label{sec:AppB}
We briefly outline how the formula (\ref{eq:SETpert}) for $\langle
T_{uu}\rangle_{conf}$ is derived from (\ref{eq:deltaW}) via
(\ref{eq:quantSETconf}).

$\delta W$ involves the integral
\begin{equation}
   z_1~z_2~\int_{t_1^*}^{t_2^*} \frac{f(t^*)}{(t^*-t_0)^3}dt^*.
\label{ex:int}
\end{equation}
Because of the prefactor $z_1z_2$, it is convenient to change the variable of
integration from $t$ (the coordinate of a point $p$ on the profile of
$\Sigma_0$ in the $z-t$ plane) to $z$ (the coordinate of a point $p$ on the
line $v=v_0$ joining $p_1$ and $p_2$, and having the same retarded time,
$u=u^*$). Since $t^*=\frac{1}{2}(u^*+v^*)=\frac{1}{2}(u-b^2/u)$ for a point
$p^*$ on $\Sigma_0$ and $z=\frac{1}{2}(v_0 - u)$, the formal transformation is
\begin{equation}
   t^* - t_0 = -\frac{z}{uv_0}(b^2+uv_0), ~~~~~t_0 =
      \frac{1}{2v_0}(v_0^2-b^2).
\label{eq:ttouv}
\end{equation}
We thus find
\begin{equation}
   \frac{1}{2}f(t^*)\frac{dt^*}{(t^*-t_0)^3}=A(z)dz,~~~~~
      A(z)=\frac{mq}{[z(z+s)]^3}~f(t^*),
\label{eq:Adef}
\end{equation}
where $m$, $q$ and $s$ are defined in (\ref{eq:notdefs}), and here $v=v_0$.

Writing, for any function $F(z)$,
\settoheight{\parht}{${\frac{1}{2}}$}
\settodepth{\pardp}{${\frac{1}{2}}$}
\begin{equation}
   \bar{F}=\frac{1}{2}\left\{\rule[\pardp]{0cm}{\parht}F(z_1)+F(z_2)\right\},
      ~~~~~ \Delta F = F(z_2)-F(z_1),
\label{eq:meandiffdef}
\end{equation}
the expression (\ref{eq:deltaW}) for $\delta W$ now reduces to 
\begin{equation}
   \frac{\pi^2v_0(\Delta z)^2}{(z_1+s)(z_2+s)}\delta W(p_1,p_2)=
      \frac{z_1z_2}{\Delta z}\int_{z_1}^{z_2}A(z)dz - \overline{z^2A(z)}.
\label{eq:Wpert2}
\end{equation}

It is now straightforward, though tedious, to carry out the operations in
(\ref{eq:quantSETconf}). The following general identities are of help in this
regard. Define, for an arbitrary $F(z)$,
\begin{equation}
   \epsilon = \frac{1}{\Delta z} \int_{z_1}^{z_2} F(z)dz - \bar{F}.
\label{eq:Fdefn}
\end{equation}
Then
\begin{eqnarray}
   \frac{\epsilon}{(\Delta z)^2} &=& -\frac{1}{12}~\overline{F''} +
      \frac{1}{120}~(\Delta z)^2~\overline{F^{(4)}}+\ldots
\label{eq:ident1}\\
   \frac{\Delta F}{\Delta z} &=& \overline{F'}-\frac{1}{12}~(\Delta z)^2~
      \overline{F'''}+\frac{1}{120}~(\Delta z)^4~\overline{F^{(5)}}+\ldots
\label{eq:ident2} 
\end{eqnarray}
where primes and subscripts in parentheses denote derivatives with respect to
$z$. If
\begin{equation}
   J = \frac{1}{(\Delta z)^2}\left\{\frac{z_1z_2}{\Delta z} \int_{z_1}^{z_2} 
      F(z)dz - \overline{z^2 F} \right\}
\label{eq:Jdef}
\end{equation}
then
\begin{equation}
   \frac{\di^2 J}{\di z_1 z_2} = - 2 \frac{\epsilon}{(\Delta z)^2}
      \left(1+\frac{6z_1z_2}{(\Delta z)^2}\right)+\frac{2}{(\Delta z)^2}
      +\left(\frac{\Delta(zF)}{\Delta z}-\bar{F}
      -\frac{1}{2}\frac{\Delta(z^2 F')}{\Delta z}\right).
\label{eq:ident3}
\end{equation}
Taking coincidence limits $z_2 \rightarrow z_1$ gives
\begin{eqnarray}
   \left[\frac{\di^2 J}{\di z_1 \di z_2}\right] &=& 
      - \frac{1}{60} (z^2F^{(4)} + 10zF'''+20F''),
\label{eq:ident4}\\
\settoheight{\parht}{${\displaystyle \frac{\di J}{\di z}}$}
\settodepth{\pardp}{${\displaystyle \frac{\di J}{\di z \di z}}$}
   \left[\rule[0.85\pardp]{0cm}{0.85\parht}\overline{\frac{\di J}{\di z}}\right] &=& 
      - \frac{1}{24}(z^2F'''+8zF'' +12F'),
\label{eq:ident5}\\
   \left[J\right] &=& - \frac{1}{12}(z^2F''+6(zF)'),
\label{eq:ident6}\\
\settoheight{\parht}{${\displaystyle \frac{\di^2 J}{\di z^2}}$}
\settodepth{\pardp}{${\displaystyle \frac{\di^2 J}{\di z^2}}$}
   \left[\rule[0.85\pardp]{0cm}{0.85\parht}
      \overline{\frac{\di^2 J}{\di z^2}}\right] &=& \frac{3}{2}\left[
      \frac{\di^2 J}{\di z_1 \di z_2}\right].
\label{eq:ident7}
\end{eqnarray}

A fairly long calculation then leads to the result (\ref{eq:SETpert}) for the
conformal flux. For the canonical flux, we find
\begin{equation}
   \delta\langle T_{uu}\rangle_{min} = \frac{1}{240 \pi^2 v r} 
      \left\{\frac{q}{m^3n}~\frac{d^2 \alpha}{d \chi^2} 
      + a_1 \frac{d \alpha}{d \chi} + a_2 ~\alpha 
      +a_3~\frac{1}{m}\left(f-\frac{df}{d\chi}\right) + a_4~m f\right\}
\label{eq:SETpertmin}
\end{equation}
as the perturbation of the uniform acceleration result (\ref{eq:constSETmin}).
Here,
\begin{eqnarray}
   a_1 &=& \frac{1}{2mn}+\frac{1}{mr}+\frac{r}{m^3n^2}\left(\frac{7}{2}m^2 -
      n^2-\frac{5}{2}mn\right),
\label{eq:a1}\\
   a_2 &=& -\frac{1}{2mn}+\frac{3}{2} \frac{1}{n^2} + \frac{3}{2}\frac{1}{nr}
      +\frac{1}{r^2} + \frac{16r}{n^3}-\frac{16r}{mn^2}-\frac{1}{mr},
\label{eq:a2}\\
   a_3 &=& - \frac{1}{2n}+\frac{3}{2}\frac{m}{nr}-2\frac{m^2}{nr^2}-\frac{1}{r}
      +\frac{m}{r^2}+\frac{3}{2}\frac{m}{n^2}-\frac{m^2}{n^2r}
\nonumber\\
   &&~~~~~~~~~~ -\frac{2m^2}{n^3} +\frac{16r(m-n)}{n^3}-\frac{45m(m-n)r}{n^4},
\label{eq:a3}\\
   a_4 &=& \frac{2}{n^3}+\frac{1}{n^2r}+\frac{2}{nr^2}+\frac{60(m-n)r}{n^5}.
\label{eq:a4}
\end{eqnarray}

\section{Evaluation of $T_{vv}$}\label{sec:AppC}

For arbitrary points $p_1$ and $p_2$ in the $z-t$ plane, the expression
(\ref{eq:deltaW}) for $\delta W$ can be recast in the form
\begin{eqnarray}
   -\frac{2\pi^2}{\Delta u}&(\bar{v}\Delta u - \bar{u} \Delta v)^3 &\delta
      W(p_1,p_2) 
\nonumber \\
   &&= (R_1^2-b^2)(R_2^2-b^2)\left\{\frac{z_1z_2}{\Delta z}
      \int_{z_1}^{z_2} H(z)dz-\overline{z^2 H}\right\} 
\nonumber \\
   &&~~~~ + \frac{1}{2} \Delta v
      \left\{(u_1u_2\bar{v}+b^2\bar{u})\Delta(z^2H)-(u_1u_2\Delta v - 
      b^2 \Delta u)\overline{z^2H}\right\},
\label{eq:Wpert3}
\end{eqnarray}
which generalizes (\ref{eq:Wpert2}) to the case where $\Delta v = v_2 - v_1
\neq 0$. The integral of $H(z)$ is to be taken along the straight-line segment
joining $p_1$ and $p_2$, and we have defined 
\begin{equation}
   H(z) = \frac{u(u^2+b^2)}{[u(t^*-t_0)]^3}f(t^*).
\label{eq:Hdefn}
\end{equation}
The denominator involves a quadratic function of retarded time, having the
explicit form
\begin{equation}
   u(t^*-t_0) = \frac{1}{2}(u^2 -c u - b^2),
\end{equation}
where 
\begin{equation}
   c=\frac{(\bar{v}^2-b^2)\Delta u-(\bar{u}^2-b^2)\Delta v 
      - \frac{1}{2}\Delta u \Delta v \Delta z}{\bar{v}\Delta u-\bar{u}
      \Delta v}.
\end{equation}

By assigning different values to the ratio $\Delta v / \Delta u$ as $\Delta u
\rightarrow 0$, we approach the coincidence limit $p_1=p_2$ in all possible
directions in the $z-t$ plane. (This is subject to the restriction that $t_0$,
given by (\ref{eq:zetadef}), should be kept outside the interval
$(t_1^*,t_2^*)$ to keep the integral of $H$ in (\ref{eq:Wpert3})
mathematically well defined.) Both of the components $T_{uu}$ and $T_{vv}$
can thus be found from the corresponding directional derivatives.

We calculated $T_{vv}$ with the aid of the computer algebra package MAPLE.
Even so, the calculation is not straightforward. The main obstacle is the
evaluation of the integral term in (\ref{eq:Wpert3}). After taking the
appropriate $v_1$ and $v_2$ derivatives of (\ref{eq:Wpert3}), and then the
partial coincidence limit $v_2\rightarrow v_1$, we express the integrand of
the first term in a Laurent series in $(u_2-u_1)$ and $(u-u_1)$, both to fifth
order. The result is expressed in ratios of powers of $(u-u_1)$ and
$(u_2-u_1)$, with coefficients of these ratios being functions of $u_1$ and
$v_1$ alone. The integration is then trivially performed. The first two terms
in the integrated Laurent series diverge in the coincidence limit $u_2
\rightarrow u_1$. However, these are exactly cancelled by terms arising from
the $v$ derivatives of the second term of (\ref{eq:Wpert3}).  Taking the
(trivial) coincidence limit of the remaining terms we find the following
expression for $T_{vv}$,
\begin{equation}
   T_{vv}=-\frac{1}{45 \pi^2}\frac{q^2}{v^3n^3r^3m}\left(q(\alpha+f)
      +p\frac{df}{d\chi}\right),
\label{eq:Cvv}
\end{equation}
where our notation is defined in (\ref{eq:notdefs}). We have verified that 
the expressions (\ref{eq:Cvv}) for $T_{vv}$ and (\ref{eq:SETpert}) for 
$T_{uu}$ satisfy the conservation identity (\ref{eq:consEq}).

\end{document}